\documentclass{czjphys}
\usepackage{epsfig}
\begin{document}
\title{Charge transport through a flexible molecular junction%
\footnote{
 This paper is dedicated to Ji\v{r}\'{\i} Hor\'{a}\v{c}ek our friend 
 and scientific mentor
 of one of us (M\v{C}) on the occasion of his sixtieth birthday}}
\authori{Martin \v{C}\'{\i}\v{z}ek\footnote{E-mail address {\tt cizek@mbox.troja.mff.cuni.cz}}}
\addressi{Institute of Theoretical Physics, Charles University, Prague, Czech Republic}  
\authorii{Michael Thoss, Wolfgang Domcke}
\addressii{Department of Chemistry,Technical University of Munich, D-85747 Garching, Germany}
\authoriii{}    \addressiii{}
\authoriv{}     \addressiv{}
\authorv{}      \addressv{}
\authorvi{}     \addressvi{}
%
\headauthor{M. \v{C}\'{\i}\v{z}ek, M. Thoss, W. Domcke}            
\headtitle{Charge transport through a flexible molecular junction} 
\lastevenhead{M. \v{C}\'{\i}\v{z}ek, M. Thoss, W. Domcke: Charge transport through a flexible molecular junction.} 

%
\refnum{A}
\daterec{XXX}    
\issuenumber{0}  \year{2005}
\setcounter{page}{1}
\maketitle

\begin{abstract}
Vibrationally inelastic electron transport through a flexible molecular
junction is investigated. The study is based on a mechanistic model 
for a biphenyl molecule between two metal electrodes. 
Employing methods from electron-molecule scattering theory, which allow a numerically exact treatment,
we study the effect of vibrational excitation on the transmission probability for different 
parameter regimes. 
The current-voltage characteristic is analyzed for different temperatures, based  on
 a Landauer-type formula.
Furthermore, the process of electron assisted tunneling between adjacent wells 
in the torsional potential of the  molecule is discussed and the validity of approximate methods to 
describe the transmission probability is investigated.
\end{abstract}

\section{Introduction}

The field of molecular electronics, which may represent the ultimate limit of the miniaturization 
of electronic devices, has received much interest in the last decade 
(see for example \cite{hry02,nr03} and references therein). 
Although the basic idea to use molecules as active elements in electronic circuits is 
not new \cite{ar74}, only recently  
it became possible to study the conduction of single molecules experimentally
(\cite{jgsc95,rzmbt97,snulhr02,reichert02}).
In contrast to metal or semiconductor devices, in molecular electronics 
the complex structure of a single molecule
may be utilized to obtain the 
functionality of the device. For example, switching behavior and negative 
differential conductivity \cite{srim02,cr02} have recently been demonstrated experimentally
and are believed to be a consequence of geometry changes in the conducting 
molecule. 

The tremendous experimental progress has stimulated great interest 
in the theory and {\em ab initio} 
modeling of charge transport through single molecules
(see for example \cite{hry02,nr03} and references therein).
In particular, non-equilibrium Green's function methods 
in combination with  state-of-the art electronic structure
calculations have
been employed to study the conductivity of metal-molecule-metal 
junctions with fixed nuclear geometry. The dependence of the conductive properties
on the binding geometry of the molecular bridge has also been investigated \cite{ek03,zdczrk04}.

The role of the 
vibrational degrees of freedom of the molecular bridge in the conduction process and and their effect on the
functionality of the device are less well understood.
The "static"  influence of  the internal vibrational modes  
has been studied by averaging the elastic transmittance over the probability 
distribution of the vibrational degrees of freedom \cite{omwkrlm98,s03,trn03}.
The dynamical impact of the vibrational degrees of freedom on the  
tunneling current in molecular junctions has been investigated within, 
e.g., nearest neighbor tight-binding models \cite{ek00,w03,bs01,mpt03,leh04}. 
These studies have demonstrated that the vibrational
motion of the molecular bridge may result in additional (vibrational) resonance structures in 
the transmission probability which can alter the current-voltage characteristic significantly.
Dynamical effects of nuclear motion on the conductivity have also been observed experimentally.
For example, in experiments on electron transport through ${\rm H}_2$ molecules between two platinum 
electrodes \cite{snulhr02} as well as  C$_{60}$ molecules connected to gold electrodes 
\cite{pplaam00}, indications for an influence of the center-of-mass motion of the respective 
molecule on the conductivity have been found. 
Moreover, the excitation of the vibrational degrees of freedom of the molecule
provides a mechanism for heating of the molecular junction and thus is a possible source 
of instability \cite{segal02,segal03}.

%
%
\begin{figure}
 \begin{center}
   \epsfxsize=0.99\textwidth \epsfbox{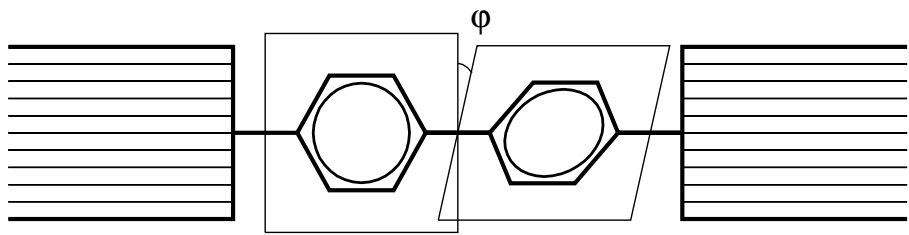}
 \end{center}
 \caption{\label{f:phi}
   Schematic illustration of the molecular junction studied. 
}
\end{figure}

Recently, we have demonstrated \cite{ctd04} 
that methods developed in the field of resonant electron-molecule
scattering \cite{d91} can be advantageously applied to study vibrationally inelastic 
effects on electron transport. In this study, the vibrational degrees of freedom
of the molecular bridge were described within the harmonic approximation.  
As an extension of this work, we investigate in this paper the influence
of large-amplitude motion of a flexible molecular bridge on charge transport through a molecular junction.
Specifically, we consider a mechanistic model for electron conduction through
a biphenyl molecule, schematically shown in Fig.~\ref{f:phi}. In biphenyl and 
similar molecules (such as bypiridine \cite{xu03} or biphenyl dithiol \cite{xr03}), 
the torsion of the two benzene rings is expected to have a strong effect
on the conductivity. Ratner, Datta and coworkers have considered this effect for a 
static nuclear bridge \cite{stdhk96} as well as in the adiabatic approximation \cite{trn03}.
{\em Ab initio} calculations of equilibrium geometries of neutral molecules
and anions for several related molecules \cite{srim02} also  indicate that the torsional
motion is presumably rather strongly coupled to the tunneling electron.
Dynamical effect of vibrations beyond the harmonic approximation have 
been studied for  similar systems based on  molecular dynamics
simulations \cite{pgdlnfs03}, where the nuclear 
motion was treated classically and the electronic structure 
was described by a density-functional
tight-binding Hamiltonian.
The purpose of this paper is to study the dynamical effect of the torsional motion on 
the transport through the bridge within a fully quantum mechanical treatment. 


This paper is organized as follows:
In section 2.1 we introduce a  generic model Hamiltonian to describe
charge transport through a  flexible molecular junction 
schematically shown in Fig.\ \ref{f:phi}. The torsional angle
between the two benzene rings is thereby treated as a dynamical variable. 
The model parameters are chosen
in accordance with available {\em ab initio} data for the biphenyl molecule.
The methods employed to calculate the inelastic transmission probability and the
current for this model are presented in section 2.2. In section 3 we discuss
the results of the model study for different parameter regimes
and analyze the performance of approximate methods, in particular
the purely elastic calculation of the transmission probability 
and the adiabatic-nuclei approximation.

\section{Theory}

\subsection{Model}

We describe charge transport through the molecular junction 
schematically depicted in Fig.\ \ref{f:phi} in the single particle approximation
as the subsequent transmission of single electrons
from a state $|\phi_k\rangle$, $k\in {\rm L}$, in the conduction band of the left lead 
into state $|\phi_k\rangle$, $k\in {\rm R}$ in the right lead. We assume
that the conduction electron can cross the junction only through a 
$\pi$-orbital $|\phi_1\rangle$ associated with the benzene ring 
which is coupled directly to the left lead and through a $\pi$-orbital $|\phi_2\rangle$
associated with the second benzene ring that is coupled to the right lead.
The torsional angle $\varphi$ of the two benzene rings is explicitly taken 
into account as a dynamical variable, since it modulates the 
overlap of the $\pi$-orbitals of the benzene rings and this will 
have an important effect on the conductivity of the molecular junction.
The influence of the other vibrational modes is neglected in the present study.
The Hamiltonian of the model thus reads
\begin{eqnarray}\label{e:ham}
 H &=& T_{\varphi}+V_0(\varphi)                      \nonumber \\
   && ~+ |\phi_1\rangle\epsilon_1\langle\phi_1|
        +|\phi_2\rangle\epsilon_2\langle\phi_2|
        +|\phi_1\rangle\beta(\varphi)\langle\phi_2|
        +|\phi_2\rangle\beta(\varphi)^*\langle\phi_1|  \nonumber \\
   && ~+ \sum_{k\in L,R}\left\{
             |\phi_k\rangle\epsilon_k\langle\phi_k|
            +\sum_{j=1,2}\left[
               |\phi_j\rangle V_k^{j}\langle\phi_k| 
              +|\phi_k\rangle V_k^{j*}\langle\phi_j|
             \right]
         \right\} 
\end{eqnarray}
The first line of Eq.\ (\ref{e:ham}) describes the torsional dynamics of the two
benzene rings in the absence of an excess electron, the second
line describes the electronic resonance states corresponding to the situation when 
the excess electron is located on the molecular bridge,
and the last line represents the free electrons in the leads and the coupling 
between the leads and the two $\pi-$orbitals. The kinetic energy operator
for the torsional motion is given by
\begin{equation}
 T_{\varphi}=-\frac{1}{2I}\frac{{\rm d}^2}{{\rm d}\varphi^2},
\end{equation}
where $I=\frac{1}{2}Md^2$ is the moment of inertia, $M$ denotes the mass of the carbon 
atom and $d$ is the diameter of the benzene
ring. The potential energy $V_0(\varphi)$ for the torsional motion 
of the isolated biphenyl molecule can be calculated
by standard methods of quantum chemistry. 
In this  study, we approximate the potential
in form of a three-term Fourier series
\begin{equation}
 V_0(\varphi)=C_0+\sum_{i=2,4,6}{\textstyle\frac{C_i}{2}}[1-\cos (i\varphi)],
\end{equation}
where the parameters $C_i$ have been adopted   from Ref.\ \cite{tum99}. We furthermore consider
a symmetric molecular junction with $\epsilon_1=\epsilon_2\equiv\epsilon_0$, 
    $V_{k\in{\rm L}}^{j=1}=V_{k\in{\rm R}}^{j=2}\equiv V_{k}$ 
and $V_{k\in{\rm L}}^{j=2}=V_{k\in{\rm R}}^{j=1}=0$.
The electronic coupling between the two rings is assumed to be of the form
\begin{equation}
 \beta(\varphi)=\beta_0 \cos(\varphi).
\end{equation}

\begin{figure}
 \begin{center}
   \epsfxsize=0.8\textwidth \epsfbox{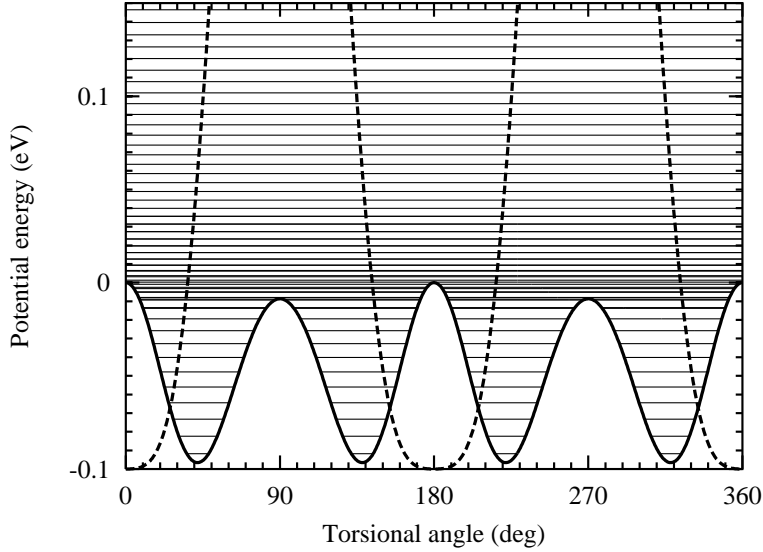}
 \end{center}
 \caption{\label{f:pot}
   Potential-energy curves for the model (for details see text). 
   The energy levels of the hindered torsional motion
   of the neutral molecule are also shown as horizontal lines. 
   Note that the states with energies well below and well above 
   zero are nearly degenerate (four-fold degenerate
   below and two-fold degenerate above zero energy, respectively).}
\end{figure}

The parameters $\epsilon_0$ and $\beta_0$ can in principle be determined from 
electronic-structure calculations of the potential-energy 
surface of the molecular anion. The vertical and adiabatic electron affinities of biphenyl have 
been calculated, e.g., in Ref. \cite{af01} yielding slightly different result for different levels of
electron structure theory.
Since the molecular electronic structure is modified by the coupling
of the biphenyl to the leads, we cannot directly adopt these data, but they can guide us to select
the parameters in a reasonable range of values.
Within our model, the adiabatic potential-energy curves of the molecular anion can be 
obtained by diagonalization of the first two lines of the Hamiltonian
(\ref{e:ham}) for  fixed nuclear geometry (i.e.\ $T_{\varphi}\approx 0$).
This way, we obtain two potential energy curves 
\begin{equation}
 V_{\rm d}^{(\pm)}(\varphi)=V_0(\varphi)+\epsilon_0\pm|\beta_0\cos(\varphi)|,
\end{equation}
corresponding to the ground and excited electronic state of the molecular anion, respectively.
To determine the parameters $\epsilon_0$ and $\beta_0$,  we impose the following requirements 
as motivated by the results of the electronic structure calculations \cite{af01} mentioned above: We
require the equilibrium geometry 
of the anion to be planar, i.e.\ $\varphi=0$, with a small, 
but positive, adiabatic electron affinity 
and, furthermore, assume a negative vertical electron affinity  of the order of -0.1eV
at the  equilibrium geometry of the neutral molecule ($\varphi_0=42^{\circ}$). 
These requirements are fulfilled for $\epsilon_0=0.9$eV and 
$\beta_0=1$eV. The corresponding potential-energy functions $V_0(\varphi)$ (full line) and 
$V_{\rm d}^{(-)}(\varphi)$ (dashed line) are shown in Fig.~\ref{f:pot}. The upper of the two 
adiabatic molecular-anion potentials, $V_{\rm d}^{(+)}(\varphi)$, which exhibits an 
avoided crossing with
$V_{\rm d}^{(-)}(\varphi)$, is too high in energy to be seen in Fig.~\ref{f:pot}.

%
%

As is well known (see for example Ref.\ \cite{d00}), the effect of the electronic coupling 
to the leads is fully described by specifying the self-energy functions 
\begin{equation}
 \Sigma_{ij}(\epsilon)
   \equiv\sum_{k}\langle\phi_i|H|\phi_k\rangle
          \frac{1}{\epsilon^{+}-\epsilon_k}
           \langle\phi_k|H|\phi_j\rangle
   =\delta_{ij}\Sigma_j(\epsilon).
\end{equation}
The presence of the factor $\delta_{ij}$ is a consequence of the fact that we 
neglect the direct coupling of the right benzene ring to the left lead and 
the coupling of the left ring to the right lead. 
As in our previous work \cite{ctd04}, we assume that the self-energies are 
proportional to the Hubbard Green's function \cite{hubbardGF} 
\begin{equation}
 \Sigma_{j}(z)=\frac{2\alpha^2}{z-\mu_j+\sqrt{(z-\mu_j)^2-4\beta^2}},
\end{equation}
which is the exact self-energy for a one dimensional semi-infinite atomic 
chain with nearest-neighbor tight-binding interaction. Within this model,
the energy of an electron in state $|\phi_k\rangle$ in lead $j$ reads
\begin{equation}
 \epsilon_k=\mu_j+2\beta\cos k.
\end{equation}
The nearest-neighbor coupling parameter
$\beta$ determines the width of the conduction band, which is $4\beta$, and $\alpha$
is the coupling strength between the last atomic site of the lead and 
the $\pi$-orbital $|\phi_j\rangle$ of the adjacent benzene ring. 
For simplicity we assume this coupling to 
be the same for both leads and independent of the angle $\varphi$. 
The chemical potential $\mu_j$ is different for both leads, the
difference $eV=\mu_{\rm L}-\mu_{\rm R}$ being the bias voltage across 
the junction. By analytic continuation we 
obtain the real and imaginary parts of the self-energy 
$\Sigma_j=\Delta_j-\frac{i}{2}\Gamma_j$ 
on the real energy axis. The imaginary part 
(which is sometimes also called 'energy-dependent decay
width of the resonance state') is given by 
\begin{equation}
 \Gamma_j(\epsilon)=\frac{\alpha^2}{\beta^2}\sqrt{4\beta^2-(\epsilon-\mu_j)^2}
\end{equation}
inside the conduction band ($|\epsilon-\mu_j|<2\beta$) and $\Gamma_j=0$ otherwise. 
In our model studies  we use a value $\beta=1$eV, which corresponds to
a conduction-band width of 4 eV. The 
strength $\alpha$ of the coupling between the molecule and the leads
is varied as described below.

\subsection{Calculation of transmission probability and current}

Employing scattering theory, the transmission
probability for  scattering of a conduction electron 
from the initial state
$|\phi_{k_{\rm i}}\rangle$ in the left lead  to the final state 
$|\phi_{k_{\rm f}}\rangle$ in the right  lead, accompanied by a vibrational transition
of the molecule from initial state $|\chi_{v_{\rm i}}\rangle$  to final state 
 $|\chi_{v_{\rm f}}\rangle$ is given by
\begin{equation}\label{e:tful}
 t_{\rm L\to R}(\epsilon_{\rm i},v_{\rm i},\epsilon_{\rm f},v_{\rm f}) = 
 \delta(E_{v_{\rm f}}+\epsilon_{\rm f}-E_{v_{\rm i}}-\epsilon_{\rm i})
 \Gamma_1(\epsilon_{\rm i})\Gamma_2(\epsilon_{\rm f})
 \left|
  \langle\chi_{v_{\rm f}}|[E-H_d-F(E)]^{-1}_{21}|\chi_{v_{\rm i}}\rangle
 \right|^2.
\end{equation}
Thereby, the Dirac delta function accounts for energy conservation
\begin{equation}
 E \equiv E_{v_{\rm i}}+\epsilon_{\rm i}=E_{v_{\rm f}}+\epsilon_{\rm f},
\end{equation}
with $\epsilon_{\rm i}$ ($\epsilon_{\rm f}$) and $E_{v_{\rm i}}$ ($E_{v_{\rm f}}$) 
being the initial (final) electron and vibrational energies, respectively.
$H_d$ is a $2\times 2$ matrix in the electronic basis 
$\{|\phi_1\rangle,|\phi_2\rangle\}$ 
given by 
\begin{equation}
 H_d=\left(
      \begin{array}{cc}
        T_{\varphi}+V_0(\varphi)+\epsilon_0 & \beta(\varphi) \\
        \beta(\varphi) & T_{\varphi}+V_0(\varphi)+\epsilon_0
      \end{array}
     \right),
\end{equation}
where each matrix element is a vibrational operator. Similarly, $F(E)$ 
is a diagonal $2\times 2$ matrix with matrix elements
\begin{equation}
 F_j(E)\equiv\Sigma_j(E-T_{\varphi}-V_0(\varphi))
       =\sum_v |\chi_v\rangle \Sigma_j(E-E_v) \langle\chi_v|.
\end{equation}
Since the operator $T_{\varphi}+V_0(\varphi)$ does not commute with $\beta(\varphi)$,
the transmission probability cannot be obtained in a closed analytical formula. However,
the expression (\ref{e:tful}) can  rather easily be evaluated numerically adopting, for example,
the free-rotor basis for the theoretical treatment of the torsional motion of the bridge.

Based on the transmission probability (\ref{e:tful}),
the current through the molecular bridge is calculated employing the generalized 
Landauer formula \cite{trn03}
\begin{eqnarray} \nonumber
 I &=& \frac{1}{\pi}\sum_{v_{\rm i},v_{\rm f}}p_{v_{\rm i}}
   \int{\rm d}\epsilon_{\rm i}\int{\rm d}\epsilon_{\rm f}
   \left\{
    f_L(\epsilon_{\rm i})[1-f_R(\epsilon_{\rm f})]
     t_{\rm L\to R}(\epsilon_{\rm i},v_{\rm i},\epsilon_{\rm f},v_{\rm f})
   \right.
\\ && \qquad\qquad\qquad\qquad\qquad
   \left.
    -f_R(\epsilon_{\rm i})[1-f_L(\epsilon_{\rm f})]
     t_{\rm R\to L}(\epsilon_{\rm i},v_{\rm i},\epsilon_{\rm f},v_{\rm f})
   \right\}.
\label{e:curr}
\end{eqnarray}
Here, 
\begin{equation}\label{e:mb}
 p_{v_{\rm i}}=\exp(-E_{v_{\rm i}}/kT)/Z
\end{equation}
denotes the Maxwell-Boltzmann distribution of the initial vibrational states
for a given temperature $T$ and 
\begin{equation}\label{e:fd}
 f_j(E)=\frac{1}{1+\exp[(E-\mu_j)/kT]}
\end{equation}
is the Fermi-Dirac distribution for the conduction-electrons in the leads. The 
bias voltage enters the formula through the chemical potentials 
$\mu_{\rm L}=eV/2$ and $\mu_{\rm R}=-eV/2$ for the left and right
leads, respectively, and is thus also present in the transmission function 
through the dependence of $\Gamma_j(\epsilon)$ on the
chemical potentials $\mu_j$.

%
%
\begin{figure}
 \begin{center}
   \epsfxsize=0.8\textwidth \epsfbox{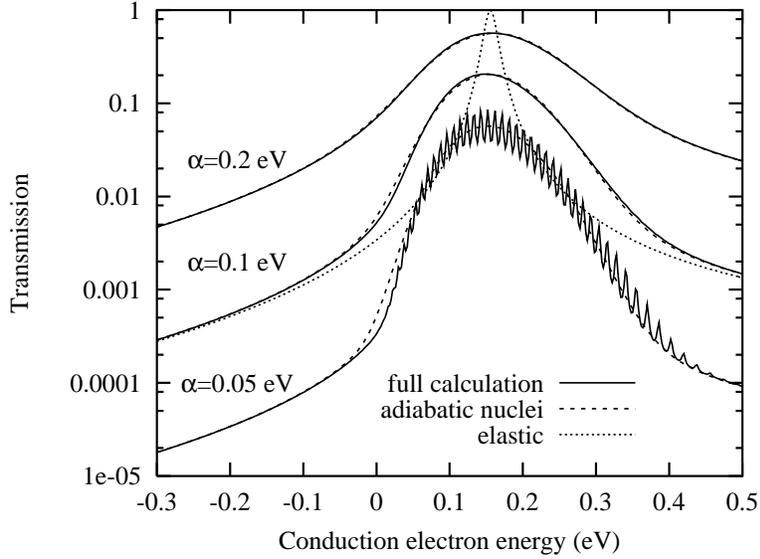}
 \end{center}
 \caption{\label{f:ttot}
   Energy dependence of the averaged transmission probability (see text) 
   for three coupling strengths $\alpha$ and temperature $T=0$. 
   The adiabatic nuclei approximation (dashed lines)
   is compared with the full nonadiabatic calculation (full lines).
   The elastic transmission calculated at the  equilibrium geometry of the neutral molecule
   is also shown for one case
   (dotted line).}
\end{figure}

\section{Results and discussion}

\subsection{Transmission probability}
The transmission probabilities 
$t_{\rm L\to R}(\epsilon_{\rm i},v_{\rm i},\epsilon_{\rm f},v_{\rm f})$
for zero bias
have been calculated according to formula (\ref{e:tful}) 
employing a free rotor basis
with typically 200 basis functions. To illustrate the results,
we average $t_{\rm L\to R}$ over the lowest four vibrational energy levels $v_{\rm i}=0,1,2,3$ 
(which  are nearly degenerate due to the symmetry of the torsional potential), 
sum over all final states $v_{\rm f}$ and integrate over
the final  energy of the electron $\epsilon_{\rm f}$. The resulting 
transmission probability is shown for three values of the coupling 
strength ($\alpha=0.05$ eV, 0.1 eV and 0.2 eV) in figure \ref{f:ttot}. 
It is seen that the transmission is essentially nonzero only in the region where the molecular 
resonance is located. 
The process can thus be understood as resonant tunneling of the electron 
from the left lead to the right 
lead through the eigenstates of $H_{\rm d}$. These states coincide (in the relevant 
energy region) with the torsional eigenstates in the potential $V_{\rm d}^{(-)}(\varphi)$
(shown as dashed line in Fig.~\ref{f:pot}). 
In the transmission probability, the individual
states are resolved only for the smallest value of the coupling $\alpha=0.05$ eV.
For larger values of $\alpha$, the individual peaks are wiped out due to the stronger 
coupling with the leads. 

It is also interesting to compare the full transmission probability 
with the elastic one, calculated for frozen vibrations
\begin{equation}
 t_{\rm L\to R}^{(\rm el)}(\epsilon_{\rm i},\epsilon_{\rm f},\varphi)=
 \delta(\epsilon_{\rm f}-\epsilon_{\rm i})
 \Gamma_1(\epsilon_{\rm i})\Gamma_2(\epsilon_{\rm f})
 \frac{\beta(\varphi)^2}{
   \left\{
     [\epsilon_{\rm i}-\epsilon_0-\Sigma_1(\epsilon_{\rm i})]
     [\epsilon_{\rm f}-\epsilon_0-\Sigma_2(\epsilon_{\rm f})]-\beta(\varphi)^2
   \right\}^2
 }.
\end{equation}
The elastic transmission function $t_{\rm L\to R}^{(\rm el)}$ depends on the 
instantaneous value of the torsional angle
$\varphi$. Fig.\ \ref{f:ttot} shows this function (dotted line) 
at the equilibrium geometry
of the neutral molecule ($\varphi_0=42^{\circ}$), integrated over the final energy
of the electron $\epsilon_f$ for 
one value of the electronic coupling ($\alpha=0.1$ eV). It is seen that 
the elastic transmission is much narrower than
the full inelastic transmission, thus demonstrating the pronounced 
broadening effect of the torsional degree of freedom. 
To investigate if this is a truly dynamical effect or rather the 
result of the distribution of torsional angles
which contribute to 
the quantum-mechanical initial state $|\chi_{v_{\rm i}}\rangle$,
we have averaged $t_{\rm L\to R}^{(\rm el)}$ over the initial distribution of the torsional 
angle $\varphi$  for temperature $T$. Integrating furthermore over the final energy 
of the electron, we obtain the expression
\begin{equation}
 t_{\rm L\to R}^{(\rm AN)}(\epsilon,\varphi)=
  \sum_v p_v \int {\rm d}\varphi \int {\rm d}\epsilon_{\rm f} |\chi_v(\varphi)|^2
  t_{\rm L\to R}^{(\rm el)}(\epsilon,\epsilon_{\rm f},\varphi),
\end{equation}
which  in the context of
electron-molecule scattering  is also
called the adiabatic-nuclei approximation \cite{l80}. As the results in  Fig.~\ref{f:ttot}
show, the adiabatic-nuclei approximation works very well
for the  system under consideration except for extremely small 
coupling to the leads where individual
vibrational states of the molecular anion are resolved, which 
is not reproduced in adiabatic-nuclei approximation.

%
%
\begin{figure}
 \begin{center}
   \epsfxsize=0.8\textwidth \epsfbox{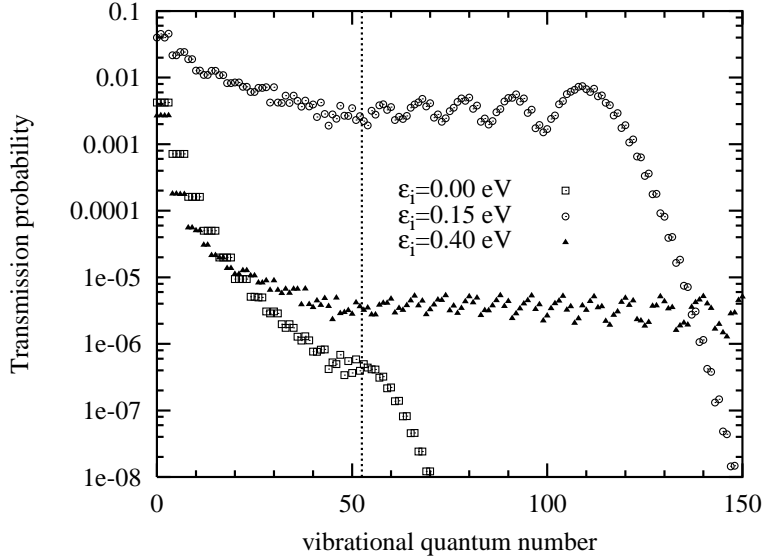}
 \end{center}
 \caption{\label{f:distra}
   Dependence of the transmission probability on the final torsional state for three 
   different initial energies of the electron: below resonance ($\epsilon_{\rm i}=0$), 
   on resonance ($\epsilon_{\rm i}=0.15$ eV) and 
   above resonance ($\epsilon_{\rm i}=0.4$ eV). 
   The vertical dotted line shows the position of the state 
   with energy $E_v=0$ (onset of freely rotating states) (cf. Fig.~\ref{f:pot}).}
\end{figure}

%
%
\begin{figure}
 \begin{center}
   \epsfxsize=0.8\textwidth \epsfbox{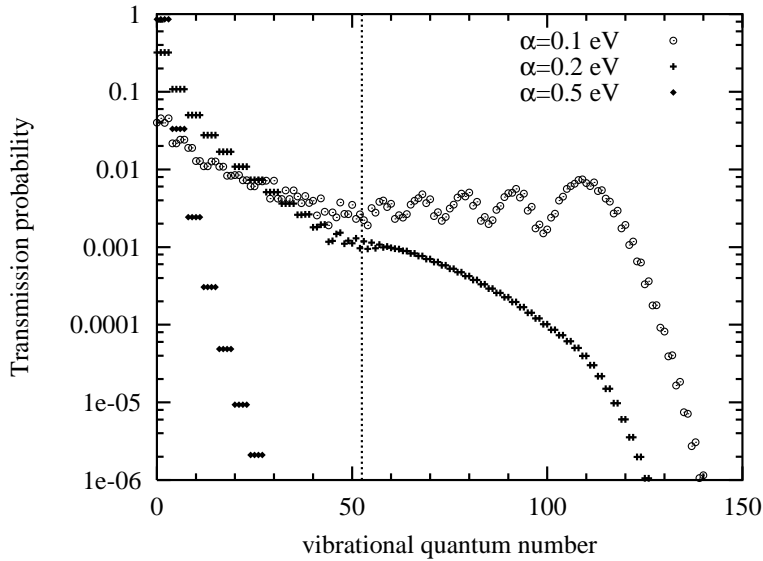}
 \end{center}
 \caption{\label{f:distrb}
   Dependence of the transmission probability on the final torsional 
   state for three different strengths of the coupling of the molecule to the leads
and an initial energy of the electron, $\epsilon_{\rm i} = 0.15$ eV.
}
\end{figure}

Being based solely on elastic transmission mechanisms, however,
the adiabatic-nuclei approximation  cannot describe the vibrational 
excitation of the molecular bridge accompanying the transmission of the electron.
This information, which is important to characterize the possible heating
of the molecular bridge, is contained in the 
full transmission probability (\ref{e:tful}). As an illustration,  Fig.~\ref{f:distra}
shows the dependence of the inelastic transmission probability (integrated over the final
electron energy) on the final torsional state for a coupling strength of  $\alpha=0.1$ eV and 
three different initial energies of the electron. 
The results exhibit a substantial vibrational excitation 
of the bridge  in the case of resonant electron transmission.
The effect of the electronic coupling strength $\alpha$  on this resonant heating process
is studied in Fig.~\ref{f:distrb}. It is seen that highly excited torsional 
states are not populated  for strong coupling to the leads (corresponding to a  broad resonance), 
although in this case the total transmission probability is larger than for small values of $\alpha$
(cf.\ Fig.\ \ref{f:ttot}).
This result is due to the fact that for strong coupling between molecular bridge and leads
the residence time of the electron on the molecular bridge 
is too short to allow efficient electron-nuclear interaction 
and thus the total transmission is dominated by the elastic contribution.


%
%
\begin{figure}
 \begin{center}
   \epsfxsize=0.8\textwidth \epsfbox{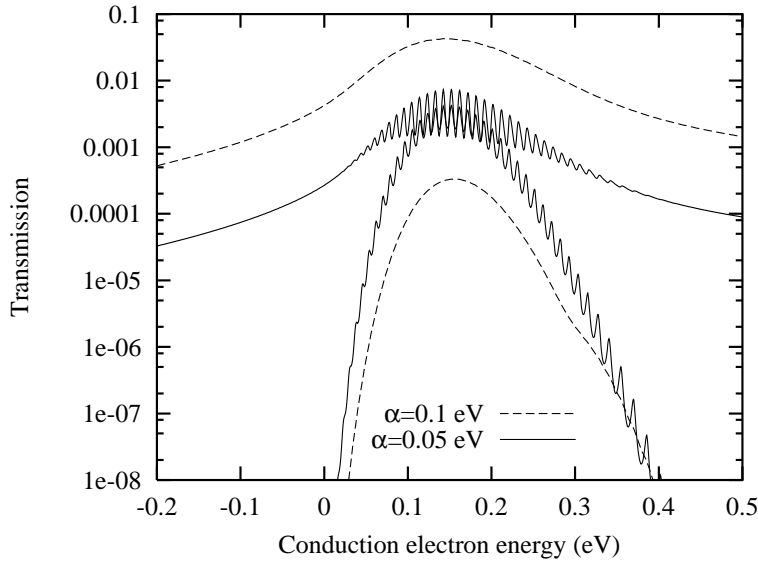}
 \end{center}
 \caption{\label{f:tswap}
   Electron-assisted tunneling between adjacent wells in the torsional potential.
   Shown is the energy dependence of the electron transmission probability
   starting from a torsional state which is localized around $\varphi_0$
   with (lower two curves) 
   and without (upper two curves) transition to another, 
   equivalent torsional state in the neighboring well.}
\end{figure}

In the context of electron transport through molecular bridges,
vibrational excitation (and possibly dissociation) 
of the molecular bridge is presumably the most important process
induced by the coupling between electron and vibrational degrees of freedom.
Another interesting process is electron assisted tunneling between the wells
of the torsional potential $V_0(\varphi)$. To study this process, we consider  torsional states
which are localized in the wells of the potential $V_0(\varphi)$. 
The potential $V_0(\varphi)$  has
four equivalent equilibrium geometries $\varphi_{0,1}=\pm\varphi_0$ and 
$\varphi_{2,3}=\pi\pm\varphi_0$
(see Fig.\ \ref{f:pot}). As a consequence, combinations of
the four nearly degenerate states with lowest energy
can be used to obtain states which are localized in the four wells. 
Due to the near-degeneracy, these states
are close to stationary states.
In Fig.\ \ref{f:tswap} we
show the transmission probability of the electron
starting from a torsional state located at $\varphi_0$ and 
ending up in the same state (upper two curves) or in a torsional state that is 
localized in the adjacent
potential well at $\varphi_1$  (lower two curves). 
It is seen that the resonance peak for both processes is located at the same position, the width
of the peak, however, is significantly narrower for the lower two curves, 
i.\ e.\ for current-induced tunneling between 
the potential wells. This can be understood as suppression of the electron transmission 
due to (assisted) tunneling through the torsional barrier. 
Another interesting result is that the  transmission probabilities for the two different processes
at energies close to the resonance peak become very similar for smaller coupling
($\alpha = 0.05$ eV, full lines). This is a consequence of the long-lived character of the molecular
anion in the case of small coupling to the leads. For sufficiently long lifetimes
of the molecular anion, the torsional state in the potential
$V_{\rm d}^{(-)}$ has the same probability to decay to both neighboring wells 
in the neutral potential $V_0$. 

%
%
\begin{figure}
 \begin{center}
   \epsfxsize=0.8\textwidth \epsfbox{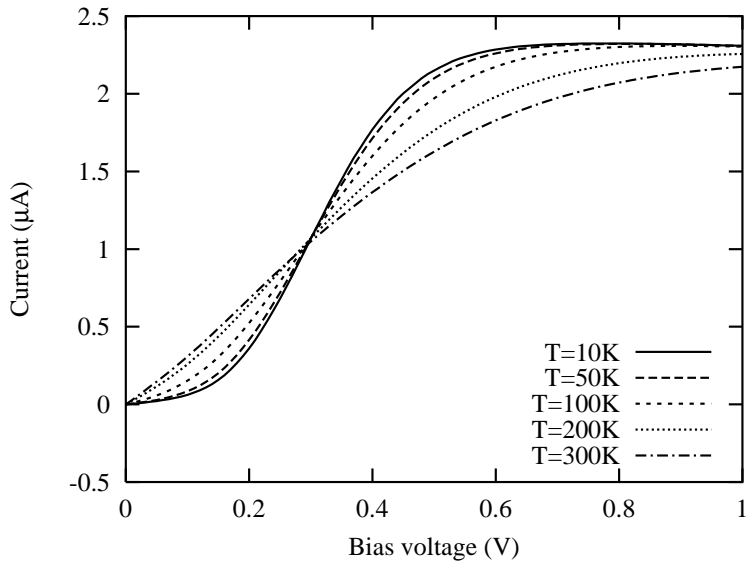}
 \end{center}
 \caption{\label{f:current}
   Current-voltage characteristics for different temperatures (see legend) and an electronic
 coupling strength  $\alpha=0.1$eV.}
\end{figure}


\subsection{Current-voltage characteristic}

Finally, we consider the current-voltage characteristic for our model.
To obtain the current for a given voltage via Eq.\ (\ref{e:curr}),
we have calculated the transmission coefficients (\ref{e:tful}) for a series of bias 
voltages on a dense energy grid.  
Fig.~\ref{f:current} shows results for different temperatures $T$.
The $I-V$ characteristics exhibit a step-like structure. Due to thermal broadening,
the width of the step increases with increasing temperature.
It should be emphasized, that this thermal broadening is mainly a result of the thermal excitation of
the torsional degree of freedom (i.e.\ Eq.\ (\ref{e:mb})), 
whereas the thermal effect of the electrons (in Eq.\ (\ref{e:fd})) is almost negligible. 

The adiabatic-nuclei approximation (data not shown) follows 
the full calculation closely. For voltages above 0.3 V, the difference is not noticeable.
It becomes more apparent close to zero bias. For example, for $T=100$ K 
the zero-bias conductivity is $4.5\times 10^{-4}$, whereas the adiabatic nuclei approximation
gives $6.7\times 10^{-4}$. This difference is a consequence of the energy loss during 
the transmission process which is correctly described in the full calculation, 
but not in the adiabatic-nuclei approximation.

\section{Conclusions}

We have analyzed vibrationally inelastic electron transport 
through a molecular junction in the presence 
of an anharmonic large-amplitude motion of the molecular bridge.
The study was based on a mechanistic model for a biphenyl molecule between two metal electrodes.
The results of the study demonstrate the effects of electron-vibrational coupling on the 
transmission probability and the current through the junction.
While for small  coupling between the molecule and the leads the
torsional degree of freedom gives rise to well resolved structures 
in the transmission function, for stronger coupling it mainly results in a broadening, which
increases significantly for higher temperature.  
We have, furthermore, studied the vibrational excitation process accompanying the transmission
of the electron through the molecular bridge, which is 
particularly pronounced for  electron energies close to the molecular resonance state and/or small
to moderate electronic coupling strength.
A correct description of the vibrational excitation process 
is also important to characterize the heating of the molecule caused by the electron transport. 
To investigate this heating process in more detail, the coupling to the other vibrational 
degrees of freedom of the molecule (which were neglected in the present study) and 
the environment has to be taken into account \cite{segal02,segal03,ctd04}.

Another interesting process, which is not related directly  to vibrational excitation
of the molecular bridge, are structural changes of the bridge, in particular electron assisted 
tunneling between  different wells of the torsional potential.
Our studies show that  electron assisted tunneling becomes particularly important when narrow
resonances are present. 

We have also investigated the validity of approximate methods, in particular
the purely elastic description of the transmission process and the adiabatic nuclei approximation.
As was found in our previous study for models with small amplitude motion \cite{ctd04}, the
purely elastic calculation typically predicts much too narrow transmission probabilities. 
The adiabatic nuclear approximation, on the other hand, which takes into account the 'static' effect
of the vibrational degrees of freedom due to their initial distribution,
can describe the average transmission probability rather well,
but it is not capable of  describing the vibrational excitation accompanying the electron transport.
Furthermore,  it is also expected to fail if the density of states
of the leads varies quickly in the energy range of interest. This aspect,
which is expected to be of particular importance if the molecular bridge is bound to 
semiconductor instead of metal electrodes will be the subject of future work.

\section*{Acknowledgements}

Support from the Alexander von Humboldt foundation and GA\v{C}R project 
No. 202/03/D112 is gratefully acknowledged.

\end{document}